\documentclass[twocolumn,pra,aps
reprint, superscriptaddress,
showpacs,
 amsmath,amssymb,
aps, pra,
]{revtex4}
\usepackage{amssymb}
\usepackage{amsmath}
\usepackage{cases}
\usepackage{stmaryrd}
\usepackage{tipa}
\usepackage[dvips]{graphicx}
\usepackage{graphicx}% Include figure files
\usepackage{dcolumn}% Align table columns on decimal point
\usepackage{bm}% bold math
\usepackage{graphicx}% Include figure files
\usepackage{dcolumn}% Align table columns on decimal point
\usepackage{bm}% bold math
\usepackage{color} % change color of text
\usepackage{CJK}

\def\la{{\langle}}
\def\ra{{\rangle}}

\newcommand{\beq}{\begin{equation}}
\newcommand{\eeq}{\end{equation}}
\newcommand{\beqa}{\begin{eqnarray}}
\newcommand{\eeqa}{\end{eqnarray}}

\begin{document}

\title{Transient particle energies in shortcuts to adiabatic expansions of harmonic traps}

\author{Yang-Yang Cui}
\affiliation{Department of Physics, Shanghai University, 200444
Shanghai, People's Republic of China}

\author{Xi Chen}
%\email[]{xchen@shu.edu.cn}
\affiliation{Department of Physics, Shanghai University, 200444
Shanghai, People's Republic of China}

\author{J. G. Muga}
\affiliation{Departamento de Qu\'{\i}mica-F\'{\i}sica, UPV-EHU, Apdo 644, 48080 Bilbao, Spain}
\affiliation{Department of Physics, Shanghai University, 200444
Shanghai, People's Republic of China}
\begin{abstract}
The expansion of a harmonic potential that holds a quantum particle
may be realized without any final particle excitation but much faster than adiabatically  via ``shortcuts to adiabaticity'' (STA).
While ideally the process time can be reduced to zero, practical limitations and constraints impose
minimal finite times for the externally controlled time-dependent frequency protocols.
We examine the role of different time-averaged
energies (total, kinetic, potential, non-adiabatic) and of the instantaneous power
in characterizing or selecting different protocols.
Specifically, we prove a virial theorem for STA processes, set minimal energies
for specific times or viceversa, and discuss their realizability by means of Dirac impulses
or otherwise.
\end{abstract}
%matter waves, atomic cooling methods
\maketitle
\section{Introduction}
The need to shorten the times of adiabatic process in different quantum or classical
systems has motivated a surge of activity
to design ``Shortcuts to adiabaticity'' (STA), namely,  fast protocols for the external parameters
controlling the system that are generically non-adiabatic but provide the same final populations than the adiabatic
dynamics, see \cite{Review13} and references therein.
%Much research has been devoted to
%analyze or improve  their stability and determine or minimize the costs implied by the time saved.
The time-dependent harmonic oscillator is a paradigmatic study case, as it
represents generically  the conditions near equilibrium and has allowed for experimental implementations
\cite{Schaff1,Schaff2},
and applications in many different classical or quantum systems     \cite{Kos09,MugaJPB2009,stat09,ChenPRL10,Masuda2010,Muga2010JPB,energy,SL10,SL11,SL11b,KosloffEPL2011,LAWuPRA2011,Campo2011,VignoloNJP2011,
Choi11,MugaPRA2012,ChoiPRA2012,Gonzalo2012,ff2012,Stefanatos2012,Zhang2012,Hanggi2013,Jarzynski2013,Yuste2013,
DavidMugaPRL2014,CampoSR2014,Palmero}. For anharmonic corrections see as well \cite{To2012,Lu2014}.
Since a basic aim of the shortcuts is to shorten the process
time $t_f$, some ``price'' in the form of high transient energies may be expected to be paid.
Several articles have studied the energies and protocol times involved: their characterization; their mutual relation,
which is not necessarily of the simple form of a time-energy uncertainty principle; and also their optimization
under different constraints and conditions \cite{energy,KosloffEPL2011,SL11}.
In this paper we build on these results and continue the analysis of the energies and times in STA processes. We do not aim at being comprehensive but at examining  points that either had not been discussed or needed some clarification. Our motivation here is of fundamental nature,
but the results will be relevant to quantify the third principle of thermodynamics \cite{Kos09,ChenPRL10,energy}, as well as the speed, efficiency limits, and costs
of quantum engines and refrigerators \cite{Kos09,KosloffEPL2011,SK,Hanggi2013,Jarzynski2013,CampoSR2014}.

After a quick review of invariant-based engineering in the Introduction,
we shall first put forward in Sec. \ref{tkv} a virial theorem for STA processes
which relates time-averaged total, kinetic and potential energies, $\overline{E}$, $\overline{K}$, $\overline{V}$.
We then discuss an idealized protocol that minimizes the time-averaged energy for a given time $t_f$, or viceversa,
by means of Dirac-delta impulses of the spring constant. It is shown that
these impulses contribute to the
total averaged energy and we specify how much. After a comparison of several protocols
in
a $(\overline{E},t_f)$ plane we study the non-adiabatic energy and its time-average in Sec. \ref{nae}.
Its lower bound for given $t_f$ is found and its implementation is discussed.
We also compare different protocols for this quantity. Finally, the instantaneous power in STA processes
is examined in Sec. \ref{power}, and we end up with a discussion.
\subsection{Invariant-based inverse engineering}
%
%
%
%
%Speeding up adiabatic process in harmonic trap has received a lot attention because its widely application to ``low" adiabatic process from laser cooling to quantum heat engine \cite{ChenMugaPRA2009,Kosloff2009,KosloffEPL2009}.
%\subsection{Invariant-based inverse engineering }
Here we consider a particle of mass $m$ in a harmonic trap, with time-dependent Hamiltonian
\begin{equation}
\label{Hamiltonian}
H(t)=\underbrace{\frac{p^2}{2m}}_{K}+\underbrace{\frac{1}{2}m\omega(t)^2q^2}_V,
\end{equation}
where $\omega(t)$ is the time-dependent angular frequency of the trap (from now on ``frequency'' will be understood as an angular frequency),  and $q$ and $p$ are position and momentum operators,
with $q$ being just a $c$-number in coordinate representation.
Lewis and Riesenfeld found the quadratic invariant \cite{Lewis1969}
$
I(t)=\pi^2/2m+m\omega_0^2q^2/b^2,
$
where $\omega_0=\omega(0)$ is the initial  frequency, $\pi=bp-m\dot{b}q$ is a momentum conjugate to $q/b$, the dots represent derivative with respect to time, and $b\equiv b(t)$ is a time-dependent, dimensionless scaling function proportional to the width of the elementary dynamical modes.
%,  as we shall see.
Since an invariant $I(t)$ must satisfy
%
%\begin{equation}
%\label{DInvariant}
$
-i[I,H]/\hbar+\partial I/\partial t=0,
$
%\end{equation}
%
so that $\la \Psi(t)|I(t)|\Psi(t)\ra$ remains constant for any $|\Psi(t)\ra$ that evolves with $H(t)$,
$b$ must obey the Ermakov equation
\begin{equation}
\label{Ermakov}
\ddot{b}+\omega^2(t)b=\frac{\omega_0^{2}}{b^3}.
\end{equation}
Any state evolving with $H(t)$ can be expanded as a superposition of
``dynamical modes'' $|\Psi(t)\rangle=\sum_n a_n |\psi_n(t)\rangle$, where  $n=0,1,2,..$, the $a_n$ are time-independent amplitudes, $|\psi_n(t)\rangle=e^{i\alpha_n(t)}|u_n(t)\rangle$,  $\alpha_n(t)=-(n+\frac{1}{2})\omega_0\int_0^{t}dt'/b^2(t')$
are Lewis-Riesenfeld phase factors, and the $|u_n(t)\rangle$ are eigenvectors of the invariant,
$I(t)|u_n(t)\rangle=\lambda_n|u_n(t)\rangle$, where $\lambda_n=(n+1/2)\hbar \omega_0$ are time-independent eigenvalues.

The  dynamical modes have the form
\beqa
\label{OneMode}
\la q|\psi_n(t)\rangle &=&\left(\frac{m\omega_0}{\pi\hbar}\right)^{\frac{1}{4}}
\frac{e^{-i(n+1/2)\int^t_0(\omega_0/b^2)dt'}}{(2^nn!b)^{1/2}}
\nonumber\\
&\times&e^{i\frac{m}{2\hbar}\Big(\frac{\dot{b}}{b}\!+\!i\frac{\omega_0}{b^2}\Big)q^2}
{\cal{H}}_n\Big[\Big(\frac{m\omega_0}{\hbar}\Big)^{\frac{1}{2}}\frac{q}{b}\Big],
\eeqa
where ${\cal H}_n$ is a Hermite polynomial.
If the commutation relations $[H(0),I(0)]=0$ and $[H(t_f),I(t_f)]=0$
are satisfied
the initial eigenstates of $H$ are mapped dynamically into final eigenstates,
avoiding final particle excitation \cite{ChenPRL10}.
These commutation relations indeed hold if the boundary conditions
\begin{equation}
\begin{split}
\label{boundaries}
&b(0)=1,\quad\ \dot{b}(0)=0,
\\
&b(t_f)=\gamma,\quad\dot{b}(t_f)=0,
\end{split}
\end{equation}
are satisfied, where $\gamma=\sqrt{\omega_0/\omega_f}$, and $\omega_f$ is the final frequency.
In addition,
\beq\label{ddot}
\ddot{b}(0)=\ddot{b}(t_f)=0
\eeq
may be imposed to make $\omega(t)$ continuous at $t=0$ and $t=t_f$.
To inverse engineer $\omega(t)$, $b(t)$ is designed first, interpolating with some convenient function between
the boundary conditions at $0$ and $t_f$, and $\omega(t)$ is deduced from  the Ermakov equation (\ref{Ermakov}) \cite{ChenPRL10}.
A simple protocol based on a quintic polynomial for $b(t)$,
where the coefficients are found from Eqs. (\ref{boundaries}) and (\ref{ddot}), is \cite{ChenPRL10}
%---------------------------------------------------------------------
\begin{equation}
\label{PolyAnsatz}
b(t)=6(\gamma-1)s^5-15(\gamma-1)s^4+10(\gamma-1)s^3+1,
\end{equation}
%---------------------------------------------------------------------
where $s=t/t_f$.
For very short times, imaginary frequencies appear for some intermediate times,
corresponding to a negative, concave-down potential.

Another simple solution is the ``bang-bang'' (stepwise constant)  form of the control function $\omega(t)$.
It results in particular from applying optimal control theory (OCT) to minimize the time for  given
constraints on the frequency \cite{Kos09,ChenPRL10,SL10,KosloffEPL2011}. Some relevant
expressions are provided in the Appendix.
As well, ``bang-singular-bang'' solutions result from minimizing the time-averaged energy
for the same constraints \cite{SL11}.
\section{Total, kinetic and potential energies\label{tkv}}
For the $n$-th dynamical mode, the
instantaneous energy, $E_n(t) \equiv\langle \psi_n|H (t)|\psi_n \rangle$, is given by
\beq
\label{energy}
E_n(t)=\frac{(2n+1)\hbar}{4\omega_0}
\left[\dot{b}^2+\omega^2(t)b^2+\frac{\omega_0^2}{b^2}\right].
\eeq
%
%The imaginary frequency can help to accelerate the evolution process by using shortcuts to adiabaticity  \cite{Chenprl10}, which results in
%instantaneous negative energy $E_n (t)$. To evaluate the energy cost from the time-averaged energy might not be accurate because of the negative part. So
We may divide the energy into kinetic and potential parts using the separation  $H=K+V$ in Eq. (\ref{Hamiltonian}). Then
$E_n(t) = K_n (t)+ V_n(t)$,
where
\beqa
K_n (t) &=& \frac{(2n+1)\hbar}{4\omega_0} \left(\dot{b}^2 +\frac{\omega_0^2}{b^2}\right) ,
\label{Kn}
\\
V_n (t)&=& \frac{(2n+1)\hbar}{4\omega_0}b^2  \omega^2(t).
\label{Vn}
\eeqa
%
%Its time average
%
%
%
%may be minimized by using the Euler-Lagrange equation to obtain the lower bound for time-averaged
%energy $\overline{E_n}$ \cite{energy}.

The mean value theorem implies  for the interval $(0,t_f)$, assuming that the boundary conditions (\ref{boundaries}) are satisfied  \cite{transp},
\beq
\dot{b}_{max}\ge \frac{\gamma-1}{t_f},\,\,\,\,\,\,\,\,
|\ddot{b}_{max}|\ge 2\frac{\gamma-1}{t_f^2}.
\eeq
The first inequality sets a lower bound for the maximum of the instantaneous
kinetic energy,
$K_{n,max}\ge  \frac{(2n+1)\hbar}{4\omega_0} \frac{(\gamma-1)^2}{t_f^2}$. In general we cannot set bounds for minimal or maximal values of $V_{n}$ unless
physically motivated bounds apply for both $b$ and $\omega$.
\subsection{Virial theorem for STA processes}

A virial-theorem relation will be proved
for STA processes that implies, remarkably, that the time-averaged energy is equipartitioned into kinetic and potential contributions, namely,  $\overline{K_n}=\overline{V_n}=\overline{E_n}/2$, where the overline denotes
a time average over $[0,t_f]$, i.e., for a generic $A_n(t)$ function,
\beq
\label{average energy}
\overline{{A}_n}=\frac{1}{t_f}\int^{t_f}_0 A_n(t)\, dt.
\eeq
The virial relation for STA processes is not at all an obvious result, as the ordinary quantum-mechanical  virial theorem applies to localized (square integrable) stationary waves and a time-independent Hamiltonian \cite{virial_Fock,virial_Slater,virial_Merz}.
Generalizations were proposed for time-dependent states and a time-independent Hamiltonian,
performing the average over an infinite time or over a period if the motion is periodic \cite{virial_Crawford,virial_Ma,virial_Muga}.
Finally,   in Ref. \cite{virial_Kobe} a virial theorem was found for a charged particle in
a time-dependent electromagnetic field.

Suppose first that the trap expansion is performed adiabatically over a very long time $t_f$ so that
the state remains in level $n$ from $t=0$ to $t_f$. The
ordinary virial theorem, i.e., the one formulated for stationary states, can then be applied.
For the harmonic oscillator in level $n$, it states
that $V_n=K_n=E_n/2$.
Along the adiabatic process the values
of the energies will change slowly in such a way that the relation is preserved.
We may thus take a time average and get $\overline{V_n}=\overline{K_n}=\overline{E_n}/2$.
For an arbitrary non-adiabatic process, however, this relation will not be true in general.

According to Eq. (\ref{energy}) and Eq. (\ref{average energy}), the time-averaged energy is
\beq
\label{E1}
\overline{E_n}=\frac{1}{t_f}\int_0^{t_f}\frac{(2n+1)\hbar}{4\omega_0}\left[\dot{b}^2+\omega^2(t)b^2+\frac{\omega_0^2}{b^2}\right]dt,
\eeq
or, using Ermakov's equation (\ref{Ermakov}),
\beq
\overline{E_n}=\frac{1}{t_f}\int_0^{t_f}\frac{(2n+1)\hbar}{4\omega_0}\left[2\frac{\omega_0^2}{b^2}+\dot{b}^2-\ddot{b}b\right]dt.
\eeq
By using the boundary conditions (\ref{boundaries}) and partial integration, we  have $\int_0^{t_f}(-\ddot{b}{b})dt=\int_0^{t_f}\dot{b}^2 dt$. Then the time-averaged energy can be rewritten as, see Eq. (\ref{Kn}),
\beq
\label{E2}
\overline{E_n}=\overline{E_{n,2}}
\equiv\frac{1}{t_f}\int_0^{t_f}\frac{(2n+1)\hbar}{2\omega_0}\left[\frac{\omega_0^2}{b^2}+\dot{b}^2\right]
%&=\frac{2}{t_f}\int_0^{t_f}\frac{(2n+1)\hbar}{4\omega_0}\left[\frac{\omega_0^2}{b^2}+\dot{b}^2\right]
%\\
=2\overline{K_n}.
\eeq
(We have introduced a specific notation, $\overline{E_{n,2}}$ for the integral in Eq. (\ref{E2})
to emphasize that $\overline{E_{n,2}}$ is only the averaged energy when the boundary conditions
are satisfied.)
  Since $\overline{E_n}=\overline{K_n}+\overline{V_n}$, we find, remarkably, the virial-theorem relation
\beq
\overline{K_n}=\overline{V_n}=\overline{E_n}/2.
\eeq
This implies a number of consequences. For example, minimizing any of the time-averaged energies
automatically minimizes the others. As well, non-trivial bounds may be set since bounds for one of the
averaged energies works for the others as well. Thus, in spite of the fact that $V_n$ may be
negative during some time interval, the time-average $\overline{V_n}$ must be positive, as $K_n\ge 0$;
even more, it will be bounded from below by a positive number as described in the next subsection.
Further applications in this work may be found in Sec. \ref{lower} and in the Appendix.
\subsection{The energy contribution from Dirac impulses\label{Diracsec}}
Using the Euler-Lagrange equation \cite{energy} to minimize
Eq. (\ref{E2})
or Optimal Control Theory
(OCT) \cite{SL11},
the following quasi-optimal function is found,
\beq
\label{LB ansatz}
b(t)=\sqrt{(B^2-\omega_0^2{t_f}^2) \left(\frac{t}{t_f}\right)^2 + 2B\left(\frac{t}{t_f}\right)+1},
\eeq
where $B=\sqrt{\omega_0^2 t_f^2 + \gamma^2}-1$ and the positive root should be taken;
this only satisfies the boundary conditions $b(0)=1$ and $b(t_f)=\gamma$ at the time limits
but not  the conditions on the derivatives. Therefore, as such it only provides
a bound for the time-averaged energy.
Substituting Eq. (\ref{LB ansatz}) into Eq. (\ref{E2}), this lower bound is
%------------------------------------------------------------
\beqa
\label{bound}
&&\overline{{E}_{nL}}=\frac{(2n+1)\hbar}{2\omega_0t_f^2}\Bigg\{\left(B^2-\omega_0^2t_f^2\right)-2\omega_0 t_f \nonumber\\
&&\times\left[\mathrm{arctanh}\left(\frac{B^2+B-\omega_0^2t_f^2}{\omega_0t_f}\right)
%\nonumber\\
%&&
-\mathrm{arctanh}\left(\frac{B}{\omega_0t_f}\right)\right]\Bigg\},
\nonumber\\
\eeqa
%Since the quasi-optimal $b(t)$ in Eq. (\ref{qo}) doesn't satisfy the set of boundary conditions in Eq. (\ref{bc}), we cannot use it directly to minimize the time-averaged energy (\ref{original expression average energy}). Inspired by \cite{SL11}, we set two frequencies with $\delta$-function form to the the extremes of the quasi-optimal $b(t)$ to frictionlessly engineer the system evolving to its adiabatic-like final sate. The frequency for each time interval is
%
which becomes  $\overline{E_{nL}}\approx\frac{(2n+1)\hbar}{2\omega_f t_f^2}$ for $t_f\ll1/\sqrt{\omega_0\omega_f}$ and $\gamma\gg 1$.
In \cite{energy}, polynomial ``caps'' were added around a central time-segment
defined by Eq. (\ref{LB ansatz}), to match this function with the right
boundary conditions. A numerical example for  $t_f\ll 1/\sqrt{\omega_0\omega_f}$ and $\gamma\gg1$
showed that when the cap duration $\tau$
went to zero, $\overline{E_n}\to \overline{E_{nL}}$. However the caps did contribute to the integral, actually
half of the total time-averaged energy came from them, but the significance of this
fact was not discussed.

In \cite{SL11} an alternative to the polynomial caps was put forward, namely, Dirac-delta impulses of
the control function $\omega^2(t)$ that switch the derivatives $\dot{b}(0^+)$ and $\dot{b}(t_f^-)$
to $\dot{b}(0^{-})=\dot{b}(t_f^{+})=0$,
\beq
\label{delta frequency}
\omega^2(t) =
\left\{\begin{array}{llll}
\omega_0^2, & t \leq 0^-
\\
D_0 \delta(t), & 0^- < t < 0^+
\\
\frac{\omega_0^2}{b^4}-\frac{\ddot{b}}{b}, & 0^+ \leq t \leq t_f^-
\\
D_f \delta(t), & t_f^- < t < t_f^+
\\
\omega_f^2, & t_f^+ \leq t
\end{array}\right..
\eeq
Substituting the $\delta$-terms in the Ermakov equation (\ref{Ermakov}), integrating $\ddot{b}$ around (in the immediate neighborhood of) $t=0$,
and $t_f$, and taking into account the boundary conditions in (\ref{boundaries}), $D_0$ and $D_f$ are
found to be
\beq
\label{a0af}
%\begin{split}
D_0=-\frac{\dot{b}(0^+)}{b(0)},
D_f=\frac{\dot{b}(t_f^-)}{b(t_f)}.
%\end{split}
\eeq
(The first impulse always corresponds to a negative Delta, while the second may have the two signs \cite{SL11}.)
The protocol (\ref{delta frequency}) with Dirac impulses provides a formally elegant
proof that the bound can indeed be reached, at least
in principle, because, as the boundary conditions are satisfied,  Eq. (\ref{E2}) may be used to get
Eq. (\ref{bound}) as the actual time-averaged energy of the process.
Note that these protocols minimize the time-averaged energy for a given time $t_f$, but also
minimize the time $t_f$ for a given time-averaged energy.  In the idealized processes that realize the bound the instantaneous potential energy jumps to (plus or minus) infinity due to the Dirac pulses.
It might seem that the deltas do not contribute to the averaged energy since
$\omega$ does not appear explicitly in Eq. (\ref{E2}).
However they do. To see why and how much,
%first notice that for the central
%time segment from $0+$ to $t_f-$ characterized by  Eq. (\ref{LB ansatz}), Eq. (\ref{E2})
%alone does not generally provide the time-averaged energy
% the derivatives are not satisfied.
%Instead one must use Eq. (\ref{E1}), which, integrating by parts and retaining the boundary
%terms, may be written as
%
%
%
%
%For the protocol doesn't satisfy the first and second derivative boundary conditions in Eq. (\ref{boundaries}), inspired by \cite{SL11}, we can set two frequencies with $\delta$-function form to the the extremes of this kind of protocols to frictionlessly engineer the system evolving to its adiabatic-like final sate. The frequency for each time interval is
%
%where for the time interval $t\in[0^+, t_f^-]$, the frequency can be obtained from Ermakov equation with specified $b$.
%where $\dot{b}(0^+)$ and $\dot{b}(t_f^-)$ is from the derivative of the protocol $b$ with respect to time in the time interval $t\in[0^+, t_f^1]$.
%Depending on the energy contribution from different frequencies (\ref{delta frequency}),
the time-averaged energy given by the original expression, Eq. (\ref{E1}), where no partial integration has been carried
out,
can be separated into two parts,
one from the deltas and one from the central time interval,
\beq
\label{Entot}
\overline{E_n}=\Delta_{\delta}+\overline{E_n}[0^+,t_f^-],
\eeq
where
\beqa
\label{delta}
&&\Delta_{\delta}=\int_{0^-}^{0^+}E_n(t)\,dt+\int_{t_f^-}^{t_f^+}E_n(t)\,dt,
\\
%\beq
\label{En1}
&&\overline{E_n}[0^+,t^-_f]=\int_{0^+}^{t_f^-}E_n(t)\,dt,
\eeqa
and $E_n(t)$ is given in Eq. (\ref{energy}) with the frequencies in Eq. (\ref{delta frequency}).
%Substituting Eq. (\ref{energy}) in Eq. (\ref{delta}) with corresponding frequencies (see Eq. (\ref{delta frequency}) and Eq. (\ref{a0af})), and using partial integration and the Ermakov equation (\ref{Ermakov}),
Making use of the coefficients (\ref{a0af}), the contribution from the Dirac impulses
to the time-averaged energy is
\beq
\label{delta energy}
\Delta_{\delta}=\frac{(2n+1)\hbar}{4\omega_0 t_f}\left[\dot{b}(t_f^-)b(t_f)-\dot{b}(0^+)b(0)\right].
\eeq
The other term in Eq. (\ref{Entot})  can be rewritten using partial integration as
%usin
%partial integration $-\int \ddot{b}b\,dt=\int \dot{b}^2 \,dt-\dot{b}b|_{t=0^+}^{t=t_f^-}$, $\overline{E_n^1}$ can be rewritten as
%
\beq
\label{inner}
\overline{E_n}[0^+,t_f^-]=\overline{E_{n,2}}+\Delta_{boundary},
\eeq
where
%
%\beq
%\label{En2}
%\overline{E_{n,2}}=\frac{(2n+1)\hbar}{2\omega_0 t_f}\int_{0^+}^{t_f^-}\left(\dot{b}^2+\frac{\omega_0^2}{b^2}\right)\,dt,
%\eeq
%
\beq
\label{triEn}
\Delta_{boundary}=-\frac{(2n+1)\hbar}{4\omega_0 t_f}\left[\dot{b}(t_f^-)b(t_f)-\dot{b}(0^+)b(0)\right].
\eeq
%
%the energy contribution from $\overline{E_2}$ is same as the lower bound energy $\overline{E_n^L}$, $\overline{E_2}=\overline{E_n^L}$.
Thus it turns out that
%From Eq. (\ref{delta energy}) and Eq. (\ref{triEn}), we  see that
%
\beq
\label{tri}
\Delta_{\delta}=-\Delta_{boundary},
\eeq
so, according to Eqs. (\ref{Entot}, \ref{inner}, \ref{tri}),
\beq
\label{En and En2}
\overline{E_n}=\overline{E_{n,2}}=\overline{E_{nL}}.
\eeq
%
%\beq
%\label{En2 and EnL}
%\overline{E_{n,2}}=\overline{E_n^L}=\overline{E_n^1}+\Delta_{\delta}.
%\eeq
%
Substituting the quasi-optimal protocol (\ref{LB ansatz}) in Eq. (\ref{delta energy}),
\beq
\Delta_{\delta}=\frac{(2n+1)\hbar}{4\omega_0t_f^2}(B^2-\omega_0^2t_f^2).
\eeq
When $t_f \ll 1/\sqrt{\omega_0\omega_f}$ and $\gamma \gg 1$,
%the lower bound for the energy contribution from $\delta$-form frequency has the following form
%
\beq
\Delta_{\delta} \approx \frac{(2n+1)\hbar}{4\omega_f t_f^2}=\overline{E_{nL}}/2,
\eeq
which agrees with the contribution of the caps found numerically in \cite{energy} as $\tau\to0$.
The reason for this result did however escape the authors of \cite{energy}.
%
%
%half of the asymptotic lower bound for $n$th time-averaged energy, $\overline{E_n^L}/2$.
% \approx \frac{(2n+1)\hbar}{2\omega_f t_f^2}$.
%
%
%
%
%
\subsection{Comparison of protocols}
\begin{figure}[t]
\begin{center}
\scalebox{0.5}[0.5]{\includegraphics{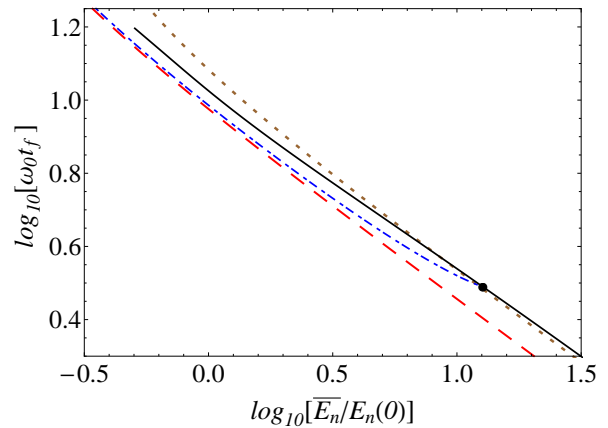}}
\caption{(Color online)
Protocol time $t_f$ versus time-averaged energy $\overline{E_n}$
%(in units of $E_n(0)$)
for a quintic polynomial protocol (dotted brown), bang-bang  for $\omega_1=\omega_2$, which decreases for larger times,
see Eq. (\ref{bbfre}) (solid black),
and lower bound (dashed red). The filled black point is for the bang-bang corresponding to
$-\omega_0^2 \leq \omega^2(t) \leq \omega_0^2$.
%,
%the bound of bang-bang control exhibited by dotted brown line is $-\omega_2^2 \leq \omega \leq \omega_2^2$, where $\omega_2 \geq \sqrt{\omega_0\omega_f}$.
The bang-singular-bang form (dot-dashed blue) is for the same bounds,
%see Eq. (\ref{BSBcontrol}),
with the bang-bang control of the  point as its limiting, minimal-time case.
Parameters: $\omega_0=2500 \times 2\pi$ Hz and $\omega_f= 25 \times2\pi$ Hz.}
\label{differentimes}
\end{center}
\end{figure}

Figure \ref{differentimes} depicts $t_f$ versus the time-averaged energy for different protocols.
A log-log representation is chosen to show the global behavior and the domains where the protocols are applicable.
As announced the fastest protocols for a given $\overline{E_n}$ are the ones that implement the
bound (\ref{bound}). Other constraints lead to different winners. For example, if $|\omega^2|$ is limited by some
predetermined value the fastest solutions are of bang-bang form \cite{ChenPRL10,KosloffEPL2011}.
If in addition to this bound the time-averaged energy is fixed, they are of bang-singular-bang form
\cite{SL11}, where the protocol in Eq. (\ref{LB ansatz}) applies in a central segment flanked by
constant-frequency intervals.
The bang-bang protocol considered in Fig. \ref{differentimes} assumes the two intermediate frequency steps
$i\omega_1$ and $\omega_1$ for durations $t_1$ and $t_2$, respectively;  $\omega_1$ decreases for
larger times, and
a maximal time (minimal $\overline{E_n}$) exists,
as discussed in the Appendix and shown in the figure.

%----------------------------------------------------------
%----------------------------------------------------------%
%
%
%
%
%
%
%
\section{Non-adiabatic energy\label{nae}}

The ``non-adiabatic energy'' is defined as the difference between the total energy and the energy of
a corresponding adiabatic process,
\beq
\begin{split}
\label{NonAdiEnergy}
E^{na}(t)&=\langle H(t)\rangle-E_{ad}(t)
\\
&=\sum_j \left[{\cal P}_j(t)-{\cal P}_j(0)\right]\epsilon_j(t),
\end{split}
\eeq
where the $\epsilon_j(t)=(j+1/2)\hbar\omega(t)$ are the instantaneous eigenenergies, the ${\cal P}_j(t)$ the corresponding
populations, and $E_{ad}\equiv\sum_j {\cal P}_j(0) \epsilon_j(t)$.
We must now assume $\omega(t)\ge 0$ in order to have a meaningful
real quantity.
%We should point out that, for the quantum adiabatic process, the system evolves slowly with constant probability $P_j(t)=P_j(0)$ and real instantaneous eigenvalues, analogous to adiabatic process, we set $\langle H_0^{a}(t)\rangle=\sum_j P_j(0)E_j(t)$ as ``adiabatic energy'' \cite{ChenMugaPRA2009} in STA process with neglecting the ``slow" property, and the angular frequency should be real for $\langle H_0(t)\rangle$ and $\langle H_0^{a}(t)\rangle$. Simultaneously, the definition of non-adiabatic energy exists when it satisfies microscopic
The ``minimal work principle'' \cite{AllahverdyanPRE2005}, establishes that
$\langle H(t) \rangle\geq E_{ad}(t)$ provided
that the initial state is passive (i.e., the initial density matrix is diagonal in the energy representation and satisfies ${\cal P}_n(0)\geq {\cal P}_{n+1}(0)$) and no level-crossings occur.
We shall restrict this section to the ground state $|\psi_0\rangle$ as a simple passive state. Then,
from Eq. (\ref{energy}),
%---------------------------------------------------------------------
\beq
\label{NonAE}
E^{na}(t)=\frac{\hbar}{4\omega_0}\left(\dot{b}^2+\omega^2(t)b^2+\frac{\omega_0^2}{b^2}\right)-\frac{1}{2}\hbar \omega(t),
\eeq
%---------------------------------------------------------------------
In the isentropic expansion stroke of an Otto cycle
the non-adiabatic energy gives the dissipated work, that must vanish at $t_f$
for a STA process. Its time average has been proposed to quantify the cost to implement a quantum engine
based on the harmonic oscillator \cite{CampoSR2014}.
%In quantum heat engine (or quantum refrigerator) with quantum otto cycle, if we consider quantum harmonic oscillator as working medium, we can use STA to implement the adiabatic expansion (or compression) steps. The state of the system is $\rho_G(\omega_0)=\frac{e^{-\beta H_0(\omega_0)}}{Z(\omega_0)}$, where $\beta$ is inverse temperature, $Z(\omega(t))\equiv Tr[e^{-\beta H_0(\omega(t))}]$ is the partition function. The dissipated work, which is equal to the difference of work distributions of adiabatic expansion process and STA expansion process, is $\delta W=\frac{\hbar}{2}[\frac{\dot{b}^2+\omega^2(t)b^2+\omega_0^2/b^2}{2\omega_0}-\omega(t)]\coth{\frac{\beta\hbar\omega_0}{2}}$. We can clearly see that, with fixed $\beta$, the non-adiabatic energy in Eq. (\ref{NonAE}) is proportional to the dissipated work $\delta W$.

The time-averaged non-adiabatic energy is
%---------------------------------------------------------------------
\beq
\label{WJHTime-averagedNonAE}
\overline{E^{na}}=\frac{\hbar\omega_0}{4t_f}\int_0^{t_f}\left(\frac{\dot{b}^2}{\omega_0^2}+\frac{\omega^2(t)b^2}{\omega_0^2}+\frac{1}{b^2}-\frac{2\omega(t)}{\omega_0}\right)dt.
\eeq
%---------------------------------------------------------------------
Using Ermakov's equation, partial integration and the conditions (\ref{boundaries}), we
get a simpler form,
%convenient to apply optical control theory,
%The frequency $\omega^2=\frac{\omega_0^2}{b^4}-\frac{\ddot{b}}{b}$ can be obtained from Ermakov equation (\ref{Ermakov}), then $\int_0^{t_f}\omega^2(t)b^2 dt=\int_0^{t_f}\frac{\omega_0^2}{b^2}-\ddot{b}b dt$. According to the boundary condition $\dot{b}(t_f)=0$ and partial integration, $\int_0^{t_f}-\ddot{b}b dt=\int_0^{t_f}\dot{b}^2 dt$. Then the time-averaged non-adiabatic energy can be simplified as
%
\beq
\label{JHTime-averagedNonAE}
\overline{E^{na}}=\overline{E_2^{na}}\equiv\frac{\hbar\omega_0}{2t_f}\int_0^{t_f}\left(\frac{\dot{b}^2}{\omega_0^2}+\frac{1}{b^2}-\frac{\omega(t)}{\omega_0}\right)dt.
\eeq
where, again, we have introduced a special notation for the last integral, $\overline{E_2^{na}}$, to emphasize that it becomes the  time average of the non-adiabatic energy
provided the boundary conditions (\ref{boundaries}) are satisfied.
%---------------------------------------------------------------------

%---------------------------------------------------------------------
%\beq
%\label{Time-averagedNonAE}
%\overline{E_0^{na}}\equiv\frac{1}{t_f}\int_0^{t_f}E_0^{na}(t)dt.
%\eeq
%---------------------------------------------------------------------
%The Mean Value T may be used to ...

%Minimizing the time-averaged non-adiabatic energy not only suppresses instantaneous excitation energy in STA process and work fluctuation of a quantum otto cycle by using STA, but also can decrease the subsequent energy cost.
%Naturally we can use optimal control theory \cite{Stefanatos2010} or generalized Euler-Lagrange theory \cite{energy} to find lower bound for $\overline{E_0^{na}}$, but the boundary conditions in Eq. (\ref{boundaries}) are not enough to find analytical solution by using both theories. In the following section, we will try to design protocols to minimize the time-averaged non-adiabatic energy from physical limit.
%
%
%
%
%
\subsection{Lower bound for time-averaged non-adiabatic energy\label{lower}}
%
%
%
%
%
%%%%%%%%%%%%%%%%%%%%%%%%%%%%%%%%%%%%%%%%%
\begin{figure}[t]
\scalebox{0.53}[0.53]{\includegraphics{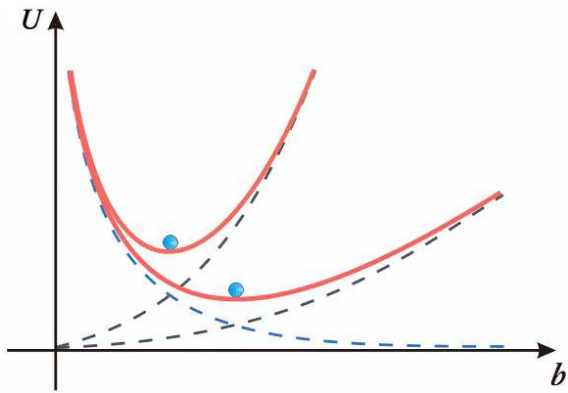}}
\caption{(Color online) A classical particle (blue dot) moves along the bottom of an expanding potential $U(t)=\left[\omega(t)^2b^2+\omega_0^2/b^2\right]/2$  (red line), see text. The two parts of $U(t)$ are
shown, $\omega(t)^2b^2/2$ (blue dashed line), and $\omega_0^2/(2b^2)$ (black dashed line).}
\label{fig1}
\end{figure}
%%%%%%%%%%%%%%%%%%%%%%%%%%%%%%%%%%%%%%%%%
A useful analogy exists between the Ermakov equation
and the dynamical equation of a fictitious (classical!) particle
with mass $m=1$ and dimensionless ``position'' $b$ moving in a potential of the form \cite{davi}
\beq
\begin{split}
{U}(t)=\frac{1}{2}\left[\omega^2(t)b^2+\omega_0^2/b^2\right]
\end{split}
\eeq
with kinetic energy $\dot{b}^2/2$.
Newton's equation takes indeed the form of the Ermakov equation,
\beq
\label{Newton}
\ddot{b}=\omega_0^2/b^3-\omega^2(t)b.
\eeq
This analogy provides useful intuition and will help us to
find the lower bound for $\overline{E^{na}}$.
%Here $b$ and $\dot{b}$ play the role of canonical coordinate and canonical momentum respectively.
The corresponding (dimensionless) energy is
\beq
\label{NHamiltonian}
{\cal H}=\frac{1}{2}\left(\dot{b}^2+\omega^2(t)b^2+\frac{\omega_0^2}{b^2}\right).
\eeq
Up to a constant factor, it has the same form than the total energy (\ref{energy}),
$E_{0}=\frac{\hbar}{2 \omega_0}{\cal H}$. Beware that the kinetic and potential energies of the classical particle
do not correspond in general to the quantum counterparts in Eqs. (\ref{Kn},\ref{Vn}).
Specifically, the term $\omega_0^2/(2b^2)$ is ``kinetic'' in the quantum scenario, and ``potential'' in the classical
analogy.
%
%where the first part in brackets represents kinetic energy.
%The Hamilton's equations are
%\beq
%\begin{split}
%\label{HEquations}
%\ddot{b}&=-\frac{\partial H}{\partial b}=-\omega(t)^2b+\frac{\omega_0^2}{b^3},\\
%\dot{b}&=\frac{\partial H}{\partial \dot{b}}.
%\end{split}
%\eeq

Let us now calculate the excitation energy of the particle $E^{ex}$ measured from the
potential minimum.
To find the minimum we calculate $d {U}(t)/db=0$, which leads to
\beq
\label{bAndOmega}
\omega(t)=\frac{\omega_0}{b^2(t)}.
\eeq
and ${U}_{min}=\omega_0\omega$, so that
\beq
\label{ex}
E^{ex}=\frac{1}{2} \dot{b}^2+\frac{1}{2}\left(\omega^2b^2+\frac{\omega_0^2}{b^2}-2\omega_0\omega\right).
\eeq
This has exactly the same form than the non-adiabatic energy (\ref{NonAE}) up to the scaling factor,
$E^{na}=\frac{\hbar}{2 \omega_0}E^{ex}$.
To complete the analogy, we consider that $\omega(t)$ changes from $\omega_0$ to $\omega_f$
and trajectories from the initial to the final points of the
potential minimum, $b(0)=1$ to $b(t_f)=\gamma$.

Note that both the first term (kinetic energy) and the second one (potential energy measured from the
minimum) are positive. The potential part of Eq. (\ref{ex}) can in fact be made zero
if the particle moves all the time at the bottom of the potential, without ever
being affected by a force, then
Eq. (\ref{bAndOmega}) should hold and, substituted in Eq. (\ref{Newton}) this gives
the equation
\beq
\ddot{b}=0,
\eeq
with solution
\beq
\label{linear}
b=1+\frac{\gamma-1}{t_f} t,
\eeq
that satisfies the boundary conditions
$b(0)=1$ and $b(t_f)=\gamma$.
This linear $b$ also minimizes the time-averaged kinetic energy
for these boundary conditions, as it can be seen from the Euler-Lagrange equation.
However the particle trajectory described by Eq. (\ref{linear}) is not at rest at $t=0$ and $t_f$,
so for the more restricted family of trajectories satisfying Eq. (\ref{boundaries})
it only provides a lower bound
for $\overline{E^{ex}}$. Of course the same (scaled) lower bound is valid for the analogous quantum system,
namely
\beq
\label{boundlinear}
\overline{E_{L}^{na}}=\frac{\hbar}{4\omega_0}\left(\frac{\gamma-1}{t_f}\right)^2.
\eeq
%
%
%For adiabatic process in quantum mechanics, the particles in harmonic potential always stay at lower excited state. In the hypothetic classical model, we suppose that this fictitious particle always moves along the bottom of the expansion trap $V(t)$ (the sketch map is in Fig. \ref{fig1}).

%----------------------------------------------------------------------------------------
%\begin{figure}
%\scalebox{0.55}[0.55]{\includegraphics{contourplotomegafvstf.eps}}
%\caption{Contour plot for the approximate lower bound of time average nonadiabatic energy in the energy unit of $\hbar\omega_0$, where $\omega_0=125\times2\pi$.}
%\label{fig.{1}}
%\end{figure}
%----------------------------------------------------------------------------------------
When $\gamma\gg1$,  $\overline{E_{L}^{na}}\approx\frac{\hbar}{4\omega_f{t_f^{2}}}$,
which is half the lower bound of the time-averaged energy for the ground dynamical mode
if, in addition, $t_f\ll1/\sqrt{\omega_0\omega_f}$. Under these conditions the
quantum kinetic energy is dominated by the first term in Eq. (\ref{Kn}), which corresponds to
the classical kinetic energy,
and the virial theorem implies $\overline{K_0}=\overline{E_0}/2$.
\subsection{Comparison of protocols}
The linear trajectory of $b(t)$ in Eq. (\ref{linear}) does not satisfy all the boundary conditions
(\ref{boundaries})
and thus the protocol $\omega(t)=\omega_0/b^2$ based on it leads to excitation.
Dirac impulses are not an option now, as the one at $t=0$ would have to be
negative but $\omega^2<0$ is not allowed.
We may instead complement the protocol with caps of durations $\tau_L$ and $\tau_S$ (for ``launching'' and ``stopping'' respectively)
that connect the linear function with the proper boundary conditions, similarly to \cite{energy}.
A hybrid protocol defined in this manner is
%----------------------------------------------------------------------------------------
\beq
\label{HybridProtocol}
 b(t) = \left\{\begin{array}{ll}
\sum_{n=0}^{3}f_n s^n, & \displaystyle{0\leq t\leq  \tau_L },
\\
(\gamma-1)s+1, & \tau_L\leq t\leq t_f-\tau_S,
\\
\sum_{n=0}^{3}g_n s^n, & t_f-\tau_S\leq t\leq t_f,
\end{array}
\right.
\eeq
%----------------------------------------------------------------------------------------
%where the first "cap" is
%----------------------------------------------------------------------------------------
%\beq
%\label{firstCap}
%\begin{split}
%b(t)& =1-\frac{t^2\left[3-3\gamma+(3t_f-4\tau_L)k\right]}{2\tau_L^2}\\
%& -\frac{t^3[-1+\gamma+(-t_f+\tau_L)k]}{\tau_L^3},
%\end{split}
%\eeq
%----------------------------------------------------------------------------------------
%the second "cap" is
%----------------------------------------------------------------------------------------
%\beq
%\label{secondCap}
%\begin{split}
%b(t)&=\gamma-\frac{3(t-t_f)^2(-1+\gamma)}{2\tau_S^2}-\frac{(t-t_f)^3(-1+\gamma)}{\tau_S^3}\\
%    &+\frac{(t-t_f)^2(-2t_f^2+2t(t_f-\tau_S)+5t_f\tau_S-4\tau_S^2)k}{2\tau_S^3},
%\end{split}
%\eeq
%----------------------------------------------------------------------------------------
where $s=t/t_f$,   the coefficients $f_n$ are found from the equations that match $b$ and $\dot{b}$ at $0$ and  $\tau_L$,
and the $g_n$ from the matching at  $t_f-\tau_S$ and $t_f$.
It can be proved that in the cubic interpolation $b> 0$.
The cap durations $\tau_{L,S}$ can be adjusted
by a subroutine that minimizes
$\overline{E^{na}}$ with the constraint $\omega\ge 0$.
Due to the constraint the cap times cannot be made zero, so, unlike Sec. \ref{Diracsec}, the bound is not reached.
The caps can only be constructed in this way until a minimal $t_f$ for which imaginary frequencies appear.
\begin{figure}[t]
\scalebox{0.55}[0.55]{\includegraphics{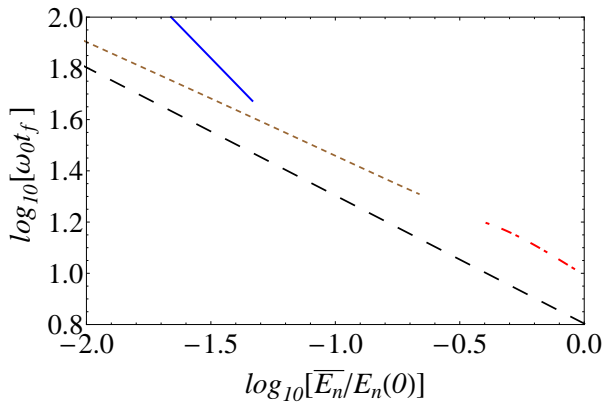}}
\caption{(Color online) Dependence of $t_f$
on the time-averaged nonadiabatic energy for the hybrid protocol in Eq. (\ref{HybridProtocol}) (dotted brown line),
quintic polynomial protocol (solid blue line), bound in Eq. (\ref{boundlinear}) (dashed black line), and bang-bang protocol with different bounds (dot-dashed red line). The parameters are $\omega_0=2500\times2\pi$ Hz and $\omega_f=25\times2\pi$ Hz.
 %The lower bound for $\overline{E_{0}^{na}}$ (dot-dashed red line), and $\overline{E_{0}}$ (dot-dashed purple line) are also given.
 }
\label{fig3}
\end{figure}%----------------------------------------------------------------------------------------
%\begin{figure}[h]
%\scalebox{0.50}[0.50]{\includegraphics{AverageNonadiabaticEnergyVStf.eps}}
%\scalebox{0.50}[0.50]{\includegraphics{AverageEnergyVStf.eps}}
%\caption{(Color online) Dependence of (a) the time-averaged nonadiabatic energy, and (b) the time-averaged energy on $t_f$ for the hybrid protocol in Eq. (\ref{HybridProtocol}) (solid blue line),
% polynomial ansatz (dashed black line), where $\omega_0=125\times2\pi$ Hz, $\omega_f=12.5\times2\pi$ Hz. The lower bound for $\overline{E_{0}^{na}}$ (dot-dashed red line), and $\overline{E_{0}}$ (dot-dashed purple line) are also given. }
%\label{fig3}
%\end{figure}
%----------------------------------------------------------------------------------------

We also consider bang-bang protocols that provide a meaningful non-adiabatic energy
with $\omega_1=0$, and $\omega_2=\omega_0 \beta$, $\beta>0$, see the Appendix,
%between $t_1$ and $t_f=t_1+t_2$.
%From the matching conditions one finds
with
\beqa
t_1&=&\frac{1}{\omega_0}\sqrt{\frac{(\gamma^2-1)(\gamma^2\beta^2-1)}{\gamma^2\beta^2}},
\\
%\eeq
%---------------------------------------------------------------------
%From Eq. (\ref{x12}) and Eq. (\ref{x1tau1}) at $t=t_1$, $t_2$ is
%---------------------------------------------------------------------
%\beq
t_2&=&\frac{1}{\omega_0\beta}\arcsin\left[\sqrt{\frac{\gamma^2-1}{\beta^2\gamma^4-1}}\right].
\eeqa

Fig. \ref{fig3}  shows the scaling of $t_f$ and the time-averaged non-adiabatic energy.
%The lower bound $\overline{E_{0L}^{na}}$ (red dash-dot line) exhibits $\overline{E_{0L}^{na}}\propto t_f^{-2}$
%behavior, and the hybrid ansatz can lower $\overline{E_{0}^{na}}$ as compared with a
%polynomial protocol.
Note that there is some complementarity among the different protocols, which are applicable in different domains.
Whereas the bang-bang protocol only satisfies the boundary conditions in a small-time, high-energy regime,
the quintic polynomial (\ref{PolyAnsatz}) or hybrid protocols apply rather in a small-energy, low-energy scenario.
\section{Power\label{power}}
%
%----------------------------------------------
\begin{figure}[t]
\scalebox{0.53}[0.53]{\includegraphics{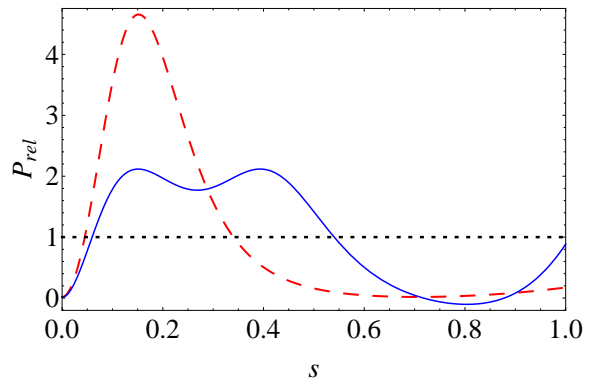}}
\caption{(Color online) Relative power versus $s=t/t_f$ for the polynomial (\ref{PolyAnsatz}) (dashed red line)
and for the polynomial (\ref{seventh}) (solid blue) with $c_3=78.5088$, $c_4=-459.7638$.
Parameters:  $\omega_0=2500\times2\pi$ Hz, $\omega_f=25\times2\pi$ Hz, and $t_f=8$ ms.
%The lower bound for $\overline{E_{0}^{na}}$ (dot-dashed red line), and $\overline{E_{0}}$ (dot-dashed purple line) are also given.
 }
\label{powerf}
\end{figure}
%-------------------------------------------------
%
We finally consider the instantaneous power during the STA process,
\beq
{P}_n(t)\equiv\frac{dE_n}{dt}=k_n\dot{\cal K}b^2,
\eeq
where we have used Eq. (\ref{energy}) and
defined $k_n\equiv\frac{\hbar(2n+1)}{4\omega_0}$ and ${\cal K}(t)\equiv \omega^2$.
This is a remarkably simple expression for $P_n$. (Defining the time-independent operator $v$ by
$V={\cal K}(t)v$, and using Eq. (\ref{Vn}), it may also be
written as $\dot{\cal{K}}\la v\ra$ which is the form given in \cite{Kos84}.)
For any STA process its integral is
\beq
\label{integ}
\int_0^{t_f} \frac{dE_n}{dt} dt=E_n(t_f)-E_n(0)=\left(n+\frac{1}{2}\right)\hbar (\omega_f-\omega_0).
\eeq

As an example of the possible use of power for shortcut design, suppose that we are interested in
minimizing the power peak (maximum) of $|{P}_n|$, distributing the power homogeneously
from 0 to $t_f$ to facilitate the power extraction.
We then set
\beq
\label{constant}
{P}_n=k_n\dot{\cal K}b^2=C_n,
\eeq
where $C_n$ is a constant
adjusted to satisfy Eq. (\ref{integ}), namely
\beq
C_n=(n+1/2)\hbar(\omega_f-\omega_0)/t_f,
\eeq
note that $C_n/k_n$ does not depend on $n$, so that the relative power
$P_{rel}\equiv P_n/C_n$ does not depend on $n$.
According to the mean value theorem, $1$ is a lower bound for the maximum of $P_{rel}$
in an arbitrary STA expansion process.
%
%Integrating  $\dot{\cal K} =C/(c_nb^2)$  we get
%
%\beq
%\int_0^{t_f} \frac{dt}{b^2}=\frac{(\omega_f+\omega_0)t_f}{2\omega_0}
%\eeq
%
%which does not depend on $n$. This is an integral constraint on $b$ that we may impose
${\cal K}$ and $b$ must satisfy the Ermakov equation (\ref{Ermakov}) so Eq. (\ref{constant}) becomes
\beq
b\dddot{b}-\ddot{b}\dot{b}+4\frac{\omega_0^2}{b^3}\dot{b}=2(\omega_0-\omega_f)\frac{\omega_0}{t_f}.
\eeq
This is a third order equation. To avoid a sudden jump in $\omega(t)$ at time $t=0$, and thus a Dirac delta
in the power,
we must impose $\ddot{b}(0)=0$ so in fact, with $b(0)=1$ and $\dot{b}(0)=0$ we already fix the three
possible constants at time $t=0$.  The numerical solution may satisfy $b(t_f)=\gamma$ for a specific $t_f$
but the conditions $\dot{b}(t_f)=\ddot{b}(t_f)=0$ fail in general.  We may thus resort to more modest objectives.
As an example of many possible strategies,
in Fig. \ref{powerf} we plot $P_{rel}$ for the fifth  order polynomial (\ref{PolyAnsatz}) and for a polynomial that satisfies
the boundary conditions (\ref{boundaries}) and (\ref{ddot}) with two free parameters that are chosen
to minimize numerically the maximum of $P_{rel}$,
\beqa
\hspace{-.3cm}&&b(t)=1\!+\!c_3s^3\!+\!c_4s^4\!-\!(21\!+\!6c_3\!+\!3c_4\!-\!21\gamma)s^5
\nonumber\\
\hspace{-.3cm}&&+(35\!+\!8c_3\!+\!3c_4\!-\!35\gamma)s^6\!-(15\!+\!3c_3\!+\!c_4\!-\!15\gamma)s^7,
\label{seventh}
\eeqa
where $s=t/t_f$.
\section{Discussion}
Different transient energies in shortcuts to adiabatic harmonic expansions
for a single particle have been studied, as well as the relations with the process time.
This is important to establish  operational limits, which are
different from the naive application of a time-energy uncertainty relation.
General bounds are provided, and protocols that realize or approach them have been discussed.
The specific experimental conditions,
i.e. the actual realization of the system and the potential by different interactions (optical, magnetic,
electrostatic),  will determine their realizability, since optimal protocols often imply Dirac deltas
(e.g. in the potential, or in the power).

The virial theorem for time averaged energies emerges as an important relation for STA processes.
We have given examples where it is instrumental in providing bounds and interpreting the results.
Further relations among various energies and their applications will be discussed elsewhere, extending the theorem to other operations and potentials.

The power in STA has been also brought to the fore. It provides a further criterion to choose
among different shortcuts. Even though its time integral is invariant for all STA, the maximal values
could be minimized.

\acknowledgments{This work was partially supported by the
NSFC (61176118 and 11474193), the Shanghai Shuguang Program (14SG35), 
the Shanghai Pujiang Program (13PJ1403000), the program for Eastern Scholar, 
the Specialized Research Fund for the Doctoral Program of Higher Education (Grant No. 2013310811003),
the Basque Government (Grant IT472-10), MINECO (Grant FIS2012-36673-C03-01), and the program UFI 11/55 of
UPV/EHU.}

\appendix
\section{Bang-bang protocols}
%
%\subsection{Time optimization: ``bang-bang" control}
%********CAREFUL WITH  THE NUMBER OF STEPS. SEE LI ET AL************
A  ``bang-bang" protocol is composed by time segments with constant frequency so that the
energy rate transfer (power) from or to the system consists of Dirac deltas at the switching times.
For applications and discussion of their relation to Optimal Control Theory see
\cite{Kos09,ChenPRL10,KosloffEPL2011,SL10,SL11}. Typically the frequency is chosen
at the extreme values allowed to achieve optimal results.

We consider here processes with two intermediate steps of the form
(more switching times may be required to find the true minimal time in general \cite{SL10})
\beq
\label{bbfre}
\omega(t)=
\left\{\begin{array}{llll}
\omega_0, & t=0
\\
i \omega_1, & 0<t < t_1
\\
\omega_2, & t_1 < t < t_1+t_2
\\
\omega_f, & t= t_f= t_1+t_2
\end{array}\right.,
\eeq
where $\omega_1\ge0$ and $\omega_2>0$ are real constants, and $t_1$ and $t_2$ are the durations
of each part.
%The bound for frequency is $-\omega_1^2 \leq \omega^2 \leq \omega_2^2$.
For $0<t<t_1$, the solution of the Ermakov equation satisfying $b(0)=1$, and $\dot{b}(0)=0$
is
\beq
\label{b1}
b(t) = \sqrt{1+ \frac{\omega^2_0 + \omega^2_1}{\omega^2_1} \sinh^2{(\omega_1 t)}},
\eeq
and for  $ t_1 \leq t \leq t_f$, the one that satisfies $b(t_f)=\gamma$, $\dot{b}(t_f)=0$ is
\beq
\label{b2}
b(t) = \sqrt{\gamma^2 + \frac{\omega^2_0- \gamma^4 \omega^2_2}{ \gamma^2 \omega^2_2} \sin^2{[\omega_2 (t_f-t)]}}.
\eeq
The matching conditions for $b(t_1)$ and $\dot{b}(t_1)$ determine
\beq
\begin{split}
\label{bang-bangt1andt2}
t_1&= \frac{1}{\omega_{1}} {\rm{arcsinh}} \sqrt{\frac{\omega_{1}^{2}(\gamma^{2}-1)(\gamma^{2}\omega_{2}^{2}-\omega_{0}^{2})}
{\gamma^{2}(\omega_{2}^{2}+\omega_{1}^{2})(\omega_{0}^{2}+\omega_{1}^{2})}},\\
t_2& =\frac{1}{\omega_{2}} \arcsin \sqrt{\frac{\omega_{2}^{2}(\gamma^{2}-1)(\gamma^{2}\omega_{1}^{2}+\omega_{0}^{2})}
{(\omega_{2}^{2}+\omega_{1}^{2})(\gamma^{4}\omega_{2}^{2}-\omega_{0}^{2})}}.
\end{split}
\eeq
There is an upper bound for $t_f$.
According to Eq. (\ref{bang-bangt1andt2}), $\omega_2 \geq \sqrt{\omega_0 \omega_f}$ must be satisfied to make $t_1 \geq 0$. With this condition, $t_2$ will be positive. At $\omega_2=\sqrt{\omega_0 \omega_f}$, $t_1=0$, which gives the upper upper bound for $t_f$ and the corresponding minimal time-averaged energy,
\beq
\begin{split}
\label{bbMaxtfandMinE}
t_f^{\textmd{max}}&=\frac{\pi}{2\sqrt{\omega_0\omega_f}},
\\
\overline{E_n^{\textmd{min}}}&=(2n+1)\hbar\frac{\omega_0+\omega_f}{4}.
\end{split}
\eeq
To calculate $\overline{E_n}$, we use the fact that the total energy remains constant
during the constant-frequency intervals. It can be calculated with
Eqs. (\ref{Kn}) and (\ref{Vn}),
at $t=0^{+}$ for the first segment and at $t=t_f^{-}$ for the second segment,
\beqa
E_n=\frac{(n+1/2)\hbar}{2}\left\{\begin{array}{ll}
(\omega_0^2-\omega_1^2)/\omega_0,& 0<t<t_1,
\\
(\omega_f^2+\omega_2^2)/\omega_f,& t_1<t<t_f.
\end{array}
\right.
\eeqa
If $\omega_1 = \omega_2 \gg \omega_0$ with $\gamma\gg 1$, $t_f$ approaches  zero, and the corresponding time-averaged energy is
\beq
\label{OPC energy bound}
\overline{E_n} \approx \frac{(2n+1)\pi \hbar}{16 \omega_f t_f^2}\ln[2 \gamma].
\eeq
This is of course
%For fixed $\omega_0$, $\omega_f$, and $t_f$, the approximate value in Eq. (\ref{OPC energy bound}) is
larger than the asymptotic value of lower bound for the time-averaged energy, $\overline{E_{nL}} \approx \frac{(2n+1) \hbar}{2 \omega_f t_f^2}$.

If we set $\omega_1=0$,
%roaching to zero, we will get the solutions of bang-bang control with $0\leq \omega \leq \omega_2$ by taking the first part of Taylor series of Eq. (\ref{b1}) and Eq. (\ref{bang-bangt1andt2}), the ansatz in $t\in[t_1,t_f]$ will be same as in Eq. (\ref{b2}). While
$\omega_2\gg \omega_0$, and $\gamma\gg 1$ then
\beq
\begin{split}
\label{bbMintfMaxE}
 t_f&\approx  t_1   \approx \frac{1}{\sqrt{\omega_0\omega_f}},
 \\
\overline{E_n} & \approx (n+1/2)\hbar\omega_0.
\end{split}
\eeq
or,
making use of the relation between $t_f$ and $\omega_0$ in Eq. (\ref{bbMintfMaxE}) we find again the
bound value, namely  $\overline{E_n}\approx \frac{(2n+1) \hbar}{2 \omega_f t_f^2}$.
We may interpret this result in view of the virial theorem. Under these conditions most of the protocol time is just a
free expansion ended up by a short segment with a strong potential. The frequency switch at time zero suddenly turns off the initial harmonic potential so that half the initial energy vanishes, and
for most of the time the energy (purely kinetic) is $E_n(0)/2$. Thus $\overline{K_n}\approx E_n(0)/2$,
and according to the virial theorem the last segment must be such that $\overline{V_n}=E_n(0)/2$
as well. That explains why $\overline{E_n}\approx E_n(0)$.
%

%----------------------------------------------------------------

\end{document}